\documentstyle[prl,aps,multicol,psfig]{revtex}
\begin{document}
\draft \wideabs{
\title{NQR of barium in BaBiO$_3$ and BaPbO$_3$}
\author{M. M. Savosta, V. D. Doroshev, V. A. Borodin, Yu. G. Pashkevich,
V. I. Kamenev and T. N. Tarasenko }
\address{Donetsk Institute of Physics \& Technics, National Academy of
Sciences of Ukraine, 83114 Donetsk, Ukraine}
\author{J. Englich and J. Kohout}
\address{Faculty of Mathematics \& Physics, Charles University,
V. Hole\v{s}ovi\v{c}k\'ach 2, 180 00 Praha 8, Czech Republic}
\author{A. G. Soldatov, S. N. Barilo and S. V. Shiryaev}
\address{Institute of Physics of Solids \& Semiconductors,
Academy of Sciences, 220072 Minsk, Belarus}
\date{\today}
\maketitle

\begin{abstract}
The NQR on Ba nuclei was studied in four samples of BaBiO$_3$ prepared
in different ways and, in addition, in BaPbO$_3$. The spectrum
of $^{137}$Ba at $T=$ 4.2 K consists of relatively broad line centered
near 18 MHz for all BaBiO$_3$ samples and near 13 MHz for BaPbO$_3$.
The integrated intensity of $^{137}$Ba resonance
in ceramic sample BaBiO$_3$ synthesized at 800$^{\circ}$C
is approximately two times larger  than in ceramic samples and single
crystal prepared at 930-1080$^{\circ}$C. The decrease of the NQR signal
can be attributed to the partial disordering of charge
disproportionated Bi ions on the two inequivalent sites.
The broadening of the resonance indicates that local distortions of
crystal structure exist in both compounds.
The point charge model was used to analyse the electric field
gradient on the Ba sites.
\end{abstract}

 }

\narrowtext

\section{Introduction}
The valence state of Bi in BaBiO$_{3}$ has long been the matter of
dispute. Two models invoked are the domination of the Bi$^{4+}$
valency implied by the composition, and the charge
disproportionation
of Bi$^{4+}$ ions into Bi$^{3+}$ and Bi$^{5+}$ (in the ionic
limit) giving rise to commensurate charge-density waves.
The latter model is based on the results of neutron diffraction
which demonstrate that the Bi ions occupy two crystallographically
inequivalent sites with significantly different Bi-O bond lengths
\cite{cox,thornton,pei}. These results are supported by EXAFS
and optical  studies \cite{balzarotti,tajima}.
On the other hand, band structure calculations \cite{mattheiss} and
defect
simulation studies \cite{zhang}, supported by the results of x-ray
photoemission \cite{nagoshi} and x-ray absorption near-edge structure
spectroscopy \cite{akhtar}, point to a minimal charge transfer between
the two Bi sites.

Using the neutron diffraction technique, Chaillout {\it et al.}
\cite{chaillout} found two types of crystal structure in BaBiO$_3$
samples prepared in different ways. The samples prepared or subsequently
heat-treated at 800$^{\circ}$C (group I) are characterized by
significant
differences between two average Bi-O distance whereas the samples
synthesized from the melt ($T = 1125^{\circ}$C, group II) are
characterized
by negligible differences. The  DTA measurements revealed an
additional phase transition at 860$^{\circ}$C on heating and
801$^{\circ}$C
on cooling in BaBiO$_3$. Chaillout {\it et al.} proposed that
this transition probably corresponds to the disproportionation of the Bi
into Bi$^{3+}$ and Bi$^{5+}$ cations and that below 800$^{\circ}$C two
structure arrangements of Bi ions exist. The first is characterized
by a partial (75\%) order of the Bi$^{3+}$ and Bi$^{5+}$ cations on the
two inequivalent sites (group I) while the second structure
corresponds to a total disorder of Bi$^{3+}$ and Bi$^{5+}$ cations
(group II).

Nuclear quadrupole resonance (NQR) can provide further insight in
the valence state of Bi because it probes locally a
distribution of electric field gradients (EFG) at the lattice sites.
However, up to now only $^{17}$O NMR studies \cite{reven} have been
performed. In the present paper we report the
first observation of the $^{135/137}$Ba NQR in BaBiO$_3$. The results
for
the samples which belong to group I and group II in a classification of
Chaillout {\it et al.} are presented. For comparison the
NQR of Ba in BaPbO$_3$ compound was also measured.

\section{Sample preparation and characterization}
Powder samples of BaBiO$_3$ were prepared by solid-state
reaction in air. For the sample 1 the starting materials BaCO$_3$
and Bi$_2$O$_3$ were mixed carefully and treated at 800$^{\circ}$C
for 8 h. After grinding and pressing to bars the sample was annealed
at 800$^{\circ}$C for 30 h. For the sample 2 the mixtures  were first
fired for 11 h with three intermediate grindings and pressings. The
temperature of firing was systematically increased from 800$^{\circ}$C
to 950$^{\circ}$C. The powder was then melted at 1080$^{\circ}$C
and quenched to room temperature afterwards.
Finally, sample 2 was quickly heated to 930$^{\circ}$C and
slowly (1.7$^{\circ}$ h$^{-1}$) cooled to achieve the oxygen
stoichiometry.
Sample 3 was obtained in the same way as sample 2 but without
melting. Single crystal of BaBiO$_3$ (sample 4) was prepared by melting
at 1080$^{\circ}$C a mixture of BaCO$_3$ and Bi$_2$O$_3$, enriched
with respect to Ba (2 at.\%).

Powder sample of BaPbO$_3$ was prepared by firing appropriate
mixtures of BaCO$_3$ and PbO at 850$^{\circ}$C for
18 h in air. After grinding and pressing to bars the sample was annealed
at 950$^{\circ}$C for 25 h.

\begin{figure}
\centerline{\psfig{figure=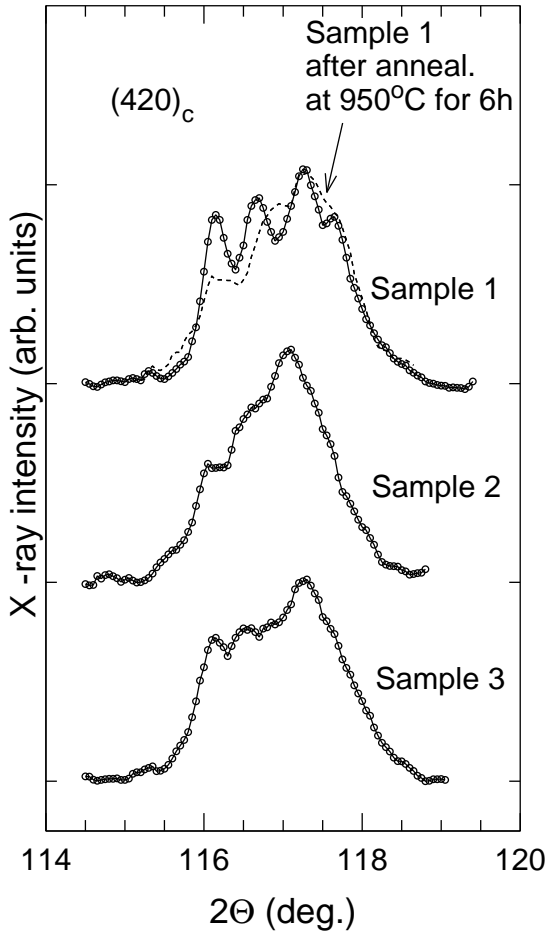,width=7.5cm,bbllx=164pt,bblly=160pt,%
bburx=422pt,bbury=598pt}} \caption{ \label{fig. 1}
X-ray powder diffraction patterns around the (420) cubic
reflection (Ni K$\alpha$ radiation) in different BaBiO$_3$ samples
prepared as described in the text. Dashed curve corresponds to the
pattern for sample 1 after it was annealed at 950$^{\circ}$C for
6h.}
\end{figure}

To characterize the samples,  x-ray powder diffraction
measurements using filtered Ni K$_\alpha$ radiation in the
2$\theta$ range 10-150$^{\circ}$ were performed at room
temperature. All compounds were found to be single phase and no
traces of starting materials were observed in the diffraction
patterns. In accordance with early reports \cite{cox,thornton,pei}
the crystal structure of BaBiO$_3$ samples was found to be
monoclinic, space group {\it I2/m}.

\begin{table}
\caption{\label{Table 1}Lattice parameters for BaBiO$_3$ samples}
\begin{tabular}{lccccc}
No & T$_{prep}$ ($^{\circ}$C) & a ($\AA$) & b ($\AA$) & c ($\AA$)
& $\beta$ \\ \hline 1 & 800 & 6.182(2) & 6.136(3) & 8.663(3) &
90.17(3) \\ 2 & 1080 & 6.187(2) & 6.143(4) & 8.673(4) & 90.14(4)
\\ 3 & 930 & 6.183(3) & 6.137(4) & 8.662(4) & 90.15(5) \\ 4 & 1080
& 6.184(2) & 6.136(3) & 8.670(3) & 90.14(3) \\
\end{tabular}
\end{table}

The unit cell parameters were determined from the splitting of the
high-angle lines, corresponding to cubic (420), (332) and (442)
reflections. The results are listed in Table I. The lattice
parameters agree well with those reported in the literature
\cite{cox,thornton,pei,chaillout}. Note that it was found in Ref.
10  that the monoclinic angle $\beta$ is larger for the samples
sintered at 800$^{\circ}$C than the angle for samples sintered at
1125$^{\circ}$C. This tendency may also be seen in Table I, though
for our data the difference is in the limit of the experimental
error.

It is interesting to note that in the present x-ray studies we
found that the width of the diffraction peaks is similar in
samples 2-4, while it is substantially  narrower in sample 1.
Shown in Fig. 1 is the diffraction pattern corresponding to the (420)
cubic reflection for polycrystalline samples under study. In order
to check whether this effect is really connected with the
temperature at which the sample is prepared, we  annealed part of
the sample 1 at 950$^{\circ}$C for 6 h. In Fig. 1 the diffraction
pattern for this sample is also presented. It is clearly seen that
the diffraction peaks in the annealed sample are broader than in
the original sample 1 and they are very similar to the ones in samples
2 and 3.

Examining the crystal structure of BaPbO$_3$, we have not found
any evidence for monoclinic distortions, accordingly the
orthorombic ({\it Ibmm}) symmetry was assumed with unit cell parameters
a = 6.0204(30), b = 6.0628(15) and c= 8.5014(15)$\AA$, which agree
well with those given by Marx {\it et al}\cite{marx}.

\section {Experimental}
The $^{135/137}$Ba spin-echo NQR experiments were performed at
4.2 and 77 K on a phase-coherent spectrometer with an averaging
technique and the complex Fourier transformation. We studied the
resonance in zero external magnetic field in the frequency region
10-25 MHz. The NQR spectra were measured recording point by point
the maximal amplitude of the Fourier transform of the echo
signal varying the transmitter frequency. The accuracy of the
amplitude measurements was $\sim$10\%. The weight of the samples
we used was about 10 g. The only exception was the sample 3 with
the weight about 5 g. The natural abundances of $^{135}$Ba and
$^{137}$Ba are 6.59\% and 11.32\%, respectively. The signal to noise
ratio for the quadrupole resonance is extremely small, therefore
relatively large averaging and the Carr-Purcel pulse sequence
\cite{englich} were employed.

The NQR spectra of barium in BaBiO$_3$ (sample 1) and BaPbO$_3$
at $T=4.2$ K are displayed in Fig. 2. Two peaks in the spectrum of
BaBiO$_3$ correspond to resonance on $^{135}$Ba
and $^{137}$Ba nuclei. The ratio of frequencies,
$\nu(^{137}$Ba$)/\nu(^{135}$Ba) = 1.54, agrees well with the ratio of
the quadrupole moments $Q(^{137}$Ba$)/Q(^{135}$Ba) = 1.556.
For BaPbO$_3$ the line near 13 MHz was attributed to the  $^{137}$Ba
nuclei. This assignment was substantiated by an absence of any
resonance near 20 MHz, which is expected if the 13 MHz line would
correspond to the $^{135}$Ba. Note that NQR of $^{137}$Ba at the same
frequency region was observed recently \cite{kumagai}
in BaPb$_x$Bi$_{1-x}$O$_3$ (x= 0.91, 0.64).

\begin{figure}
\centerline{\psfig{figure=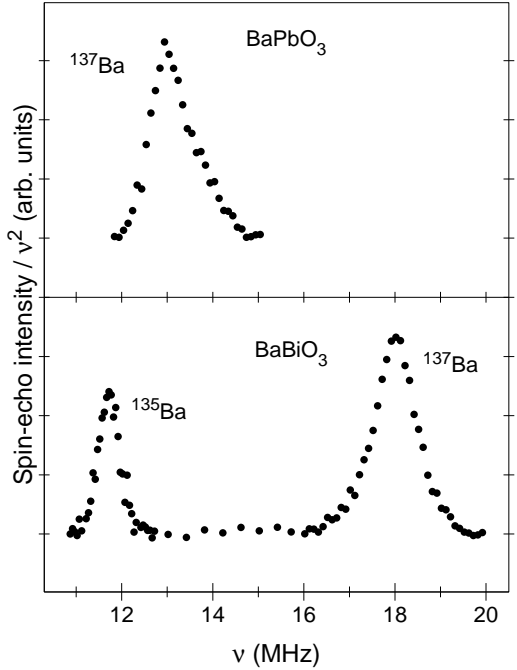,width=7cm,bbllx=120pt,bblly=163pt,%
bburx=458pt,bbury=600pt}} \caption{ \label{fig. 2} NQR spectra on
Ba nuclei in BaBiO$_3$ (sample 1) and BaPbO$_3$ taken at $T=$ 4.2
K.}

\end{figure}

The NQR spectra of $^{137}$Ba in four BaBiO$_3$ samples at $T=4.2$ K,
after the correction of the spin-echo amplitude for the weight of the
samples, are presented in Fig. 3. All spectra are centered near
18 MHz. The linewidth is $\sim$1.2 MHz for the samples 1, 2 while
it is $\sim$1.7 MHz for the samples 3 and 4.
The striking feature of the present study is that the NQR spectra
for the samples which belong to group I and group II are similar.
As will be shown below, all these spectra correspond to an ordered
arrangement of Bi$^{(4+\delta)+}$ and Bi$^{(4-\delta)+}$ cations.
Therefore the NQR data do not confirm the picture suggested by
Chaillout {\it et al.} (see Section I).
The only difference between the group I and group II,
which may be found in the spectra presented in Fig. 3, is that
the integrated intensity, which is proportional to the number of the
resonating nuclei, is approximately two times larger in the sample 1
than the  integrated intensity in the three other samples.

Nuclear spin-lattice and spin-spin relaxation times are
similar for all compounds: $T_1\simeq$3.4 s and $T_2\simeq$0.6 ms.
Therefore the relaxation can not be responsible for the
difference of the spin-echo
amplitudes for the sample 1 and samples 2-4.
In order to substantiate further the effect, the measurements
on the samples 1 and 4 at $T=77$ K were performed.
Note that due to the more suitable
set of the relaxation times ($T_1\simeq$0.04 s, $T_2\simeq$0.6
ms) we were able to use experimental conditions at which
the amplitude of the two-pulse spin echo  is practically not
influenced by the relaxation. The ratio of areas under the resonance
lines was found to be in this case $S_1/S_4\simeq$ 1.9 in a
good agreement with the data obtained at 4.2 K.

\begin{figure}
\centerline{\psfig{figure=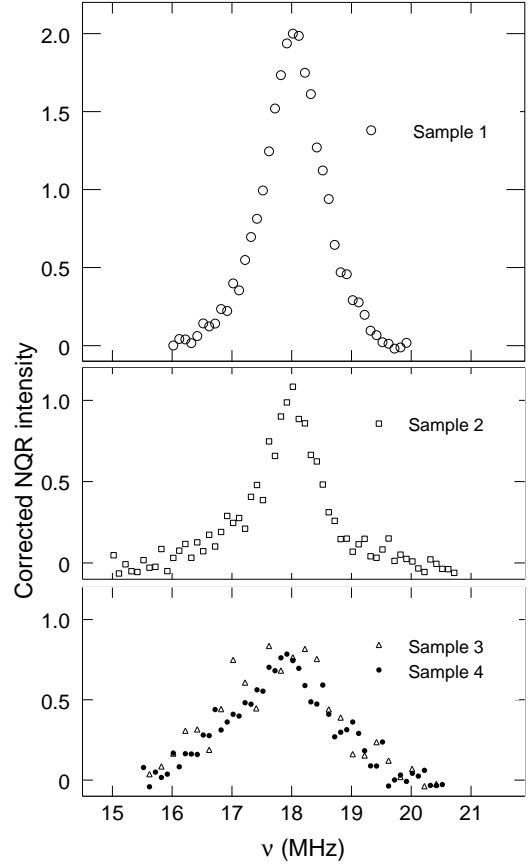,width=7cm,bbllx=100pt,bblly=70pt,%
bburx=455pt,bbury=665pt}} \caption{ $^{137}$Ba NQR spectra at $T=$
4.2 K in four  BaBiO$_3$ samples prepared in different ways. The
amplitudes are corrected for the weight of the samples. The
integrated intensity for the sample 1 is approximately two times
larger than for samples 2-4.} \label{fig. 3}
\end{figure}

\section {Discussion}
In the perovskites BaBiO$_3$ and BaPbO$_3$ barium occupies
the cavities between the Bi(Pb)-O octahedra which form
three-dimensional network. Such arrangement implies covalency in
the Bi(Pb)-O bonds, while for barium the ionic state Ba$^{2+}$
with filled electronic shells seems to be appropriate
approximation. Ionicity of the Ba$^{2+}$ bonding in the
perovskite lattice is supported by NMR studies of $^{137}$Ba in
BaTiO$_3$ \cite{bastow}, where no chemical shift within the experimental
error was found, and also by NMR in BaPb$_x$Bi$_{1-x}$O$_3$
\cite{kumagai}. This is further supported by recent
pseudopotential calculation of the electronic structure of
Ba(B,B$^{\prime}$)O$_3$ perovskites, performed by Burton and Cockayne
\cite{burton}.

Thus, when analyzing the EFG at the Ba sites, only the
contribution which arises from the charge distribution of the
surrounding ions has to be considered. In a conventional point charge
model the component $V_{ij}$ of the EFG tensor is given by
\begin{equation}
V_{ij}=(1-\gamma_{\infty})\sum_{k}e_{k}(3x_{ik}x_{jk}-
\delta_{ij}r_k^2)r_k^{-5},
\end{equation}
where $\gamma_{\infty}$ is the Sternheimer antishielding factor,
$e_k$ is the charge and $x_{ik}, x_{jk}$ are the Cartesian
coordinates of the $k$th ion with a distance $r_k$ from the
origin located at a given site.
For $^{137}$Ba (I=3/2) the NQR frequency may be written as
\begin{equation}
\nu=\frac{eQV_{zz}}{2h}\sqrt{1+\eta^2/3} ,
\end{equation}
where $Q(^{137}$Ba$)$=+0.28 $b$, $\eta=(V_{xx}-V_{yy})/V_{zz}$
is the asymmetry parameter.

In the structure of BaMO$_3$ (M=Bi, Pb) the symmetry  of
arrangement of the M ions with respect to Ba
sites is almost cubic. Consequently, if the charge distribution
is regular, the contributions of the M ions to EFG on the Ba sites
approximately cancel. This is the case for BaPbO$_3$, where the single
valence state of Pb is taken for granted, and also for BaBiO$_3$ if the
structure is ordered with respect to the disproportionated
Bi$^{(4+\delta)+}$ and Bi$^{(4-\delta)+}$ cations. Note that as
far as the $^{137}$Ba NQR is concerned,
the limiting cases of Bi$^{4+}$ cations or ordered Bi$^{3+}$
and Bi$^{5+}$ cations have nearly the same effect. The main
contribution to EFG then arises from the oxygen octahedra rotations
around the [110] cubic axes, which are known to be the main distortions
of both structures relative to an ideal perovskite crystal.

To demonstrate this point we have calculated the NQR frequency
as a function of the rotation angle $\phi_{[110]}$ for idealized
BaMO$_3$ perovskite having undistorted octahedra with a single
distance r$_{M-O}$. In the calculations the value of the
antishielding factor $\gamma_{\infty}$ = -70.7 for Ba$^{2+}$, obtained
within the framework of the local density approximation
\cite{gusev}, was used. This value agrees with the value
$\gamma_{\infty}$ = -76 mentioned in Ref. 18.
The lattice sums in (1) were calculated taking into account the
contributions from lattice sites inside a sphere  with
radius of 120$\AA$ , although the radius of 60$\AA$
was found to be large enough to obtain the EFG tensor
within the 2\% accuracy.

Shown in Fig. 4 is the calculated NQR frequency
as a function of $\phi_{[110]}$ for two values of r$_{M-O}$
corresponding to averaged Bi-O (over two sites) and Pb-O
distances. Taking the mean octahedra
tilts $\sim$8.3$^{\circ}$ for BaPbO$_3$ and $\sim$10.2-10.6$^{\circ}$
for BaBiO$_3$ ($T=$300 K)\cite{cox,thornton,pei,marx} even this
simplified model accounts for observed NQR in the two compounds.

The EFG on Ba sites in BaBiO$_3$ was calculated using crystal data
of Thornton {\it et al.}\cite{thornton} for $T=$ 4.2 and 293 K
(see Table II). In this calculation an ordered arrangement of
Bi$^{3+}$ and Bi$^{5+}$ ions was assumed. At $T=$ 4.2 K we have
obtained NQR frequency of 17.2 MHz in a good agreement to the
experimental value 18 MHz. The EFG at 293 K was found to be only
$\sim$0.85 of the value at 4.2 K reflecting the decrease of the
average octahedra tilt from $\sim$11.6$^{\circ}$ to
$\sim$10.6$^{\circ}$ in the same temperature region. The
experimentally observed decrease of the NQR frequency
$\nu$(77K)/$\nu$(4.2K) = 0.96 is in line with the above tendency.
To our knowledge no data about the crystal structure of BaPbO$_3$
at $T=$ 4.2 K exist, therefore the room temperature structure
\cite{marx} was used for the calculation. The results are given in
Table II. Taking into account that the increase of the octahedra
tilts with decreasing temperature seems to be appropriate also for
this compound \cite{kumagai}, one can conclude that point charge
model reproduces correctly the measured values of $^{137}$Ba NQR
frequency.

\begin{figure}
\centerline{\psfig{figure=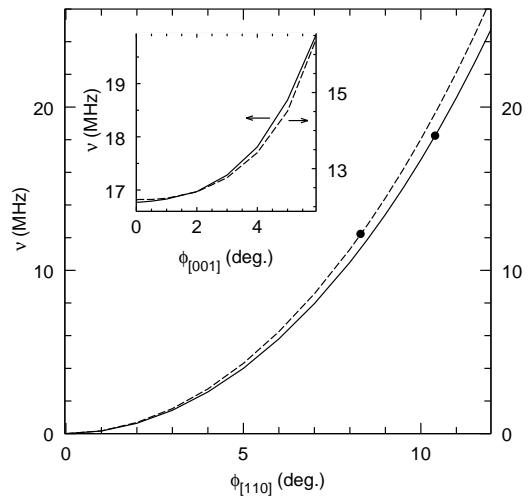,width=7cm,bbllx=90pt,bblly=184pt,%
bburx=492pt,bbury=587pt}} \caption{ The $^{137}$Ba NQR frequency
as a function of the angle of oxygen octahedra rotation around
[110] cubic axes, calculated using eqs. (1), (2). In the insert
the NQR frequency as a function of the angle of additional
octahedra rotation around [001] axes is plotted. In the
calculations the idealized perovskite BaMO$_3$ structure
consisting of  undistorted octahedra  with the single distance
r$_{M-O}$ was used. The results for r$_{M-O}$= 2.147 (dashed
curve) and 2.200 $\AA$ (solid curve) which correspond to average
Bi-O and Pb-O distances at 300 K are shown. The filled circles
correspond to the NQR frequency for $\phi_{[110]}$= 8.3$^{\circ}$
and 10.4$^{\circ}$, which are the average octahedra tilts for
BaPbO$_3$ and BaBiO$_3$ ( $T=$ 300 K), respectively.}
\label{fig.4}
\end{figure}

Let us now consider the structure of BaBiO$_3$ disordered with
respect to disproportionated Bi$^{(4+\delta)+}$ and Bi$^{(4-\delta)+}$
cations with $\delta$ substantially different from zero. In this
case a variety of random charge configurations of Bi with respect
to the Ba sites gives rise to a broad distribution of EFG over the
crystal. To simulate the situation we calculated the lattice sums
(1) for random configurations of $e_k$. The contributions from Bi
sites inside a sphere with radius of 18$\AA$ ($\sim$300 ions) were
take into account. To obtain a good statistics  $10^5$
configurations were considered. Finally, the contribution from
oxygen sites was added. The NQR spectra resulting from the above
simulation for $\delta$=0.5 and $\delta$=1 together with the one
for $\delta$=0 are plotted in Fig. 5. It is seen that the disorder
of Bi$^{(4+\delta)+}$ and Bi$^{(4-\delta)+}$ cations results in a strong
broadening of the NQR spectrum and decreases the amplitude by an
order of magnitude.

We now turn back to the experimental spectra of $^{137}$Ba in four
BaBiO$_3$ samples presented in Fig. 3. The analysis we performed
suggests that the spectra observed in all samples correspond to an
ordered arrangement of Bi$^{(4+\delta)+}$ and Bi$^{(4-\delta)+}$
cations. Decrease of the number of the nuclei experiencing this
ordered arrangement in the samples 2-4 is equivalent to the
decrease of the corresponding volume. We can conclude therefore
that samples 2-4 are {\it partially disordered} with respect to
the disproportionated Bi$^{(4+\delta)+}$ and Bi$^{(4-\delta)+}$
cations. The degree of the disproportionation $\delta$ have to be
large enough to smear out the NQR from the charge disordered
regions (see Fig. 5). In other words, the present NQR data do not
support the model in which only small if any charge transfer
between the two Bi sites exists. If the difference of Bi valencies
is small, only broadening but not a decrease of the integrated
intensity of $^{137}$Ba NQR is expected. Our conclusion in this
point  is in accord with the results of neutron diffraction
studies. However, we can't accept the model of Chaillout {\it et
al.}\cite{chaillout} (see Section I) for  group II, according to
which the structure arrangement of Bi ions corresponds to a {\it
'total disorder of Bi$^{3+}$ and Bi$^{5+}$ cations'}, as in this
case no NQR spectra at around 18 MHz are expected for the samples
2-4. The results of neutron diffraction and NQR may be
consistently explained if we assume that there exists a transition
from the long-range ordering of Bi$^{(4+\delta)+}$ and
Bi$^{(4-\delta)+}$ ($\delta\leq$1) for group I, to short-range
ordering with a coherence length smaller than the intrinsic
coherence length for neutrons (group II). Such a transition should
not lead to any substantial change in the NQR, but it may smear
out corresponding extra reflections in the neutron diffraction
patterns. The situation could be similar to the one in the
mixed-valency perovskite manganites, where inhomogeneous spatial
mixture of charge-ordered and charge-disordered microdomains with
a size of 20-30 nm was found recently \cite{mori}. It is tempting
to connect

\begin{table}
\caption{Electric field gradient on $^{137}$Ba nuclei (V$_{zz}$),
asymmetry parameter ($\eta$) and NQR frequency ($\nu_{calc}$)
calculated using the point charge model together with the
experimental values of NQR frequency ($\nu_{exp}$) for BaBiO$_3$
and BaPbO$_3$.}
\begin{tabular}{lcccccc}
compound & T(K) & V$_{zz}$(10$^{21}$V/m$^2$) & $\eta$ &
$\nu_{calc}$(MHz) & $\nu_{exp}$(MHz) \\ \hline BaBiO$_3$ & 4.2 &
-4.88 & 0.48 & 17.2 & 18.0 \\ BaBiO$_3$ & 300 & -4.15 & 0.54 &
14.7 & - \\ BaPbO$_3$ & 4.2 & - & - & - & 13.0 \\ BaPbO$_3$ & 300
& -2.80 & 0.47 & 9.84 & - \\
\end{tabular}
\label{Table 2}
\end{table}

\noindent the broadening of the x-ray diffraction peaks (Fig. 1)
for the samples 2-4 with spatial distribution of the lattice
parameters which results from such inhomogeneous mixture.

\begin{figure}
\centerline{\psfig{figure=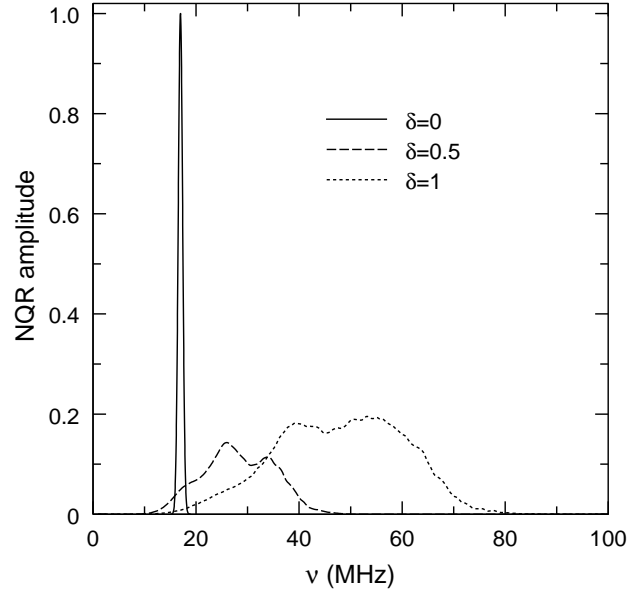,width=8cm,bbllx=90pt,bblly=191pt,%
bburx=470pt,bbury=570pt}} \caption{The simulated NQR spectra of
$^{137}$Ba in
BaBi$_{0.5}^{(4+\delta)+}$Bi$_{0.5}^{(4-\delta)+}$O$_3$ disordered
with respect to Bi ions for three values of $\delta$. The
individual contributions to the spectra are broadened by linewidth
1 MHz. The amplitudes are increased in proportion to $\nu^2$ to
reproduce the picture which should be observed experimentally.}
\label{fig. 5}
\end{figure}

Another feature of the present NQR study which was not yet
discussed is surprisingly large linewidth of NQR in both BaBiO$_3$
and BaPbO$_3$ comparing to linewidth expected for a perfectly
ordered structure. The ordinary source of the broadening, i.e. the
defects (like oxygen vacancies) or impurities is  unlikely. For
BaBiO$_3$ the samples 2 and 3 were prepared from the same mixture
of the starting materials, but they show substantially different
linewidths ($\sim$1.2 and $\sim$1.7 MHz). On the other hand, the
samples 1 and 2, produced in different ways and in different
laboratories, show identical spectra. Finally, the NQR spectrum of
BaPbO$_3$  is very similar to that obtained for
BaPb$_{0.91}$Bi$_{0.09}$O$_3$ (Ref. 14) in spite of 9\%
substitution of Bi for Pb! The strong broadening of NQR suggests
an essential local distortions of the perovskite crystal structure
for both compounds. Based on the EXAFS analysis, it was proposed
recently \cite{yacoby}, that in BaBiO$_3$ local oxygen octahedra
rotations have in addition to the [110] component a disordered
component about the [001] axis. In our idealized model of the
perovskite structure it is easy to estimate that the observed NQR
linewidth may result from a disordered rotation of
$\sim$3$^{\circ}$ about the [001] axis (see insert in Fig. 4).
Note that for the [100] and [010] axes which make the angle
45$^{\circ}$ with [110] axis, considerably smaller rotation angle
is enough to account for the NQR linewidth.

\section{Conclusions}
The NQR of barium was studied in BaBiO$_3$ and BaPbO$_3$.
The analysis of the NQR data for BaBiO$_3$ supports the double valence
of the Bi ions. Our results indicate that there exists ordering
arrangement of Bi$^{(4+\delta)+}$ and Bi$^{(4-\delta)+}$ ions
for the samples studied, irrespectively of the preparing
conditions, in contradiction with the models used until now.
We propose that the degree of the ordering is higher in the BaBiO$_3$
sample synthesized at 800$^{\circ}$C than in the ceramic samples
and single crystal, which are prepared at 930-1080$^{\circ}$C.
The broadening of the NQR suggests that the local distortions
of the structure are essential in both BaBiO$_3$ and BaPbO$_3$
compounds. The point charge model accounts for observed EFG on Ba
sites.

\acknowledgements
Our thanks to V. G. But'ko and A. A. Gusev for calculation the
antishielding factor for Ba$^{2+}$ ion.

This work was supported  in part by INTAS grant No 97-1371.

\end{document}